\documentclass[aps,prl,twocolumn,showpacs]{revtex4}
\usepackage{epsfig}
\usepackage{graphicx}
\usepackage{amsmath}

\begin{document}

\DeclareGraphicsExtensions{.eps,.EPS}

\title{A chromium dipolar Fermi sea}
\author{B. Naylor, A. Reigue, E. Mar\'echal, O. Gorceix, B. Laburthe-Tolra and L. Vernac}
\affiliation{Universit\'e Paris 13, Sorbonne Paris Cit\'e, Laboratoire de Physique des Lasers, F-93430 Villetaneuse, France
and 2 CNRS, UMR 7538, LPL, F-93430 Villetaneuse, France}

\begin{abstract}

We report on the production of a degenerate Fermi gas of $^{53}$Cr atoms, polarized in the state F=9/2, $m_F=-9/2$, by
sympathetic cooling with bosonic S=3, $m_S=-3$ $^{52}$Cr atoms. We load in an optical dipole trap 3 $\times 10^4$
$^{53}$Cr atoms with $10^6$ $^{52}$Cr atoms.
Despite this initial small number of fermionic atoms, we reach
a final temperature of $T\simeq0.6 \times T_f$ (Fermi temperature), with up to $10^3$ $^{53}$Cr atoms. This surprisingly efficient evaporation
stems from an inter-isotope scattering length $|a_{BF}| = 85 (\pm 10 )$ $a_{B}$ (Bohr radius) which is small enough to reduce evaporative
losses of the fermionic isotope, but large enough to insure thermalization.

\end{abstract}
\date{\today}
\maketitle

There has recently been tremendous activity on dipolar quantum gases. This is due to the fact that in
dipolar gases the particles interact through long-range and anisotropic dipole-dipole interactions (DDIs),
which drastically changes the nature of many-body ground and excited states. After first seminal
results with dipolar Bose-Einstein condensates made of chromium\cite{CrBEC,CrBEC2}, and then dysprosium \cite{DyBEC} and erbium \cite{ErBEC}
atoms, the focus now turns towards dipolar degenerate Fermi gases, which raise fascinating prospects
for the study of novel Fermi liquid properties \cite{BaranovZoller}, p-wave superconductivity \cite{pwave}, topological px+ipy phases \cite{pxpy},
or unconventional magnetism \cite{BuchlerPRL,Carr,BlochZoller,Buchler}.

Dipolar Fermi gases of dysprosium\cite{DyFS} and erbium\cite{ErFS} have
recently been produced, which has enabled to observe the Fermi surface deformation induced by dipolar
interactions \cite{DefFermiSurface}. However, the density of magnetically-tunable Feshbach resonances in these
Lanthanides atoms is very large \cite{FeshbachEr,FeshbachDy}, due to anisotropic short range interactions. As this density promises to be
even higher in the context of spinor gases where more than one spin state is populated, study of magnetism might be
difficult with these atoms. For this reason, chromium, with a combination of relatively
strong dipole-dipole long-range anisotropic interactions and simple isotropic short range
interactions, remains a unique atom for the study of the unconventional spinor properties of dipolar
quantum degenerate Fermi gases.

Despite its interest, producing a dipolar Fermi gas of chromium atoms has been elusive for many
years, as it represents a real experimental challenge. The main reason is the small number of
$^{53}$Cr atoms that can be captured in a magneto-optical trap (MOT), at most typically $10^5$, due to
relatively small natural abundance, the complex hyperfine structure, and, most importantly, the very large light-assisted loss rate in
a MOT \cite{MOT5253}. In this paper, we describe and demonstrate a way to produce quantum gases of fermionic
$^{53}$Cr atoms.

Our scheme consists in loading a mixture of most-abundant $^{52}$Cr atoms and
minority $^{53}$Cr atoms in the same far-detuned optical dipole trap (ODT), and in performing forced evaporative cooling.
Evaporative losses are smaller for $^{53}$Cr atoms than for $^{52}$Cr atoms, which results in very efficient evaporative
cooling characterized by a gain of typically four orders of magnitude in phase-space density for one
order of magnitude of atom losses. As a consequence, although only $3 \times 10^4$ $^{53}$Cr atoms
are loaded in the dipole trap at 60 $\mu$K before evaporation, degenerate Fermi gases of up to $10^3$ atoms
can be produced in less than 15 s. We analyze our evaporation scheme and are able to point out the decisive role played by the numerical value of
the inter-isotope scattering length, which is smaller than the bosonic scattering length. We measure this quantity, and compare it with
theoretical predictions based on mass scaling \cite{MassScaling}.

Fermi statistics leads to vanishing s-wave collisions due to Van der Walls interactions at low temperatures for polarized fermions.
Therefore, evaporative cooling to degeneracy, which requires efficient thermalization, has been achieved either with a mixture of two fermionic
Zeeman states \cite{FirstFS} for which s-wave collisions are allowed, or for a Bose-Fermi mixture \cite{Li67FS}.
With dipolar species, low temperature collisions
become possible even for identical fermions. This very peculiar effect due to the long range character of DDIs was first observed with Dy \cite{DyFS},
and was used to produce the Er Fermi sea in a very efficient way \cite{ErFS}. However for Cr the dipolar elastic cross section is 20 times smaller than
for Er, as it scales as $d^4m^2$ ($d$ being the permanent magnetic dipole, and $m$ the mass). In addition we only manage to trap
in a conservative trap about 30 times less atoms than in \cite{ErFS}, which is very unfavourable for a scenario involving only $^{53}$Cr atoms.
This is why we chose to perform sympathetic cooling with the bosonic $^{52}$Cr.

We now turn to the experimental setup used in this work with special focus on the lasers. The
strategy we used to Bose condense $^{52}$Cr is to accumulate atoms captured inside a
MOT into a superimposed far-detuned optical dipole trap (ODT) \cite{CrBECus}.
To improve loading in the ODT we make use of optical pumping from excited $^7P$ states of the MOT
trapping transition to different mestastable states, in which they are protected from
deleterious light assisted collisions \cite{MOT5253}. After the ODT is loaded, the MOT is turned off, atoms are repumped in the
$^7S_3$ ground state, optically pumped to the absolute ground state $m_s=-3$, and finally transferred
to a crossed ODT in which the evaporation is performed.

To adapt this strategy to trap both isotopes in the same ODT, new laser lines are necessary
due to the hyperfine structure of the fermionic isotope. The trapping lasers needed to obtain a $^{53}$Cr MOT
using the $^7S_3\rightarrow ^7P_4$ transition at 425 nm were reported in \cite{MOT5253}. They are generated from a dedicated laser system producing
800 mW of 425 nm light. The optimal MOT for loading the ODT contains $10^5$ $^{53}$Cr atoms, at a
temperature of 120 $\mu$K, close to the Doppler temperature. Atoms decay to metastable states $^5D_3,F=9/2$ and $^5D_4,F=11/2$ (from excited
state $^7P_4,F=11/2$), and $^5S_2,F=7/2$ (from $^7P_3,F=9/2$, see below). Repumping of these states to the ground state is performed via
the excited state $^7P_3$\cite{Note}.

The (horizontal) ODT is derived from a 100 W
Infra Red (IR) Ytterbium fibre laser at 1075 nm. We use 60 W and control the
power with an Acousto Optic Modulator (AOM). The laser is turned on and superimposed with
the MOT. Frequency modulation of the AOM at fast frequency (100 kHz) allows to adapt the IR laser
mode size with the MOT radius. We find the same AOM parameters for optimal loading of both isotopes.

\begin{figure}
\centering
\includegraphics[width= 3.0 in]{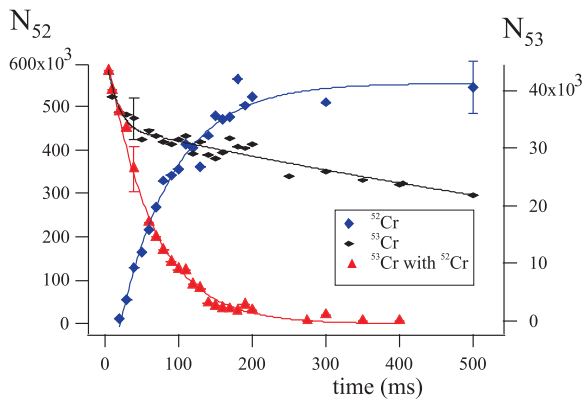}
\caption{\setlength{\baselineskip}{8pt} {\protect\scriptsize (Color online) Inelastic interspecies collisions in the IR trap. The
life time of $^{53}$Cr atoms in the ODT dramatically drops when $^{52}$Cr atoms are loaded starting
at t=0. This constrains us to optimize the Bose-Fermi mixture composition before evaporation. Solid lines are fit to the data, using exponential functions.
Error bars show statistical uncertainties.}}
\label{Loading}
\end{figure}

Optimization of the loading of the IR trap with $^{52}$Cr was studied in \cite{CrBECAppPhys}. It revealed in particular the crucial role played by
the metastable $^5S_2$ state, populated from the MOT by using a depumping beam at 427
nm (almost resonant with the $^7S_3\rightarrow ^7P_3$ transition). Optimal depumping allows to double
atom number in the ODT. For the fermionic case, we could only increase the atom number by 20 \% when depumping to $^5S_2$.
We interpret this reduction in efficiency
as a result of significantly larger inelastic collisions between $^5S_2,F=7/2$ $^{53}$Cr atoms, compared to
$^5S_2$ $^{52}$Cr atoms. This may arise from the non favorable "inverted" hyperfine structure in this state: energetically allowed two-body
inelastic collisions processes populate the other hyperfine states of $^5S_2$, leading to losses. We finally could accumulate
about $10^5$ $^{53}$Cr atoms in the ODT, to be compared with
up to $2.10^6$ for $^{52}$Cr, with respective 1/e loading time equal to 200 (50) ms.

In our experiment, the Zeeman Slower beam used to slow down $^{53}$Cr substantially reduces the $^{52}$Cr MOT atom number \cite{MOT5253}.
Therefore, we do not load the two
isotopes at the same time in the ODT. Instead we first make a $^{53}$Cr MOT (to load $^{53}$Cr atoms in the ODT), then
switch off the blue beams specific for the $^{53}$Cr MOT and turn on the $^{52}$Cr MOT for a given amount of time
$\Delta t$ (to load $^{52}$Cr atoms in the ODT). We then detect both atom numbers, $N_{52}$ and $N_{53}$. As shown
in Fig. \ref{Loading}, the presence of both isotopes in the IR trap leads to losses, attributed to inter-isotope inelastic losses \cite{Note}.
The starting point for evaporative cooling is therefore a trade-off: when $\Delta t$ increases, $N_{52}$
increases, but $N_{53}$ decreases. With $\Delta t=90$ ms, we obtain the following optimal mixture for evaporation: $N_{53}=3.10^4$ and
$N_{52}=1.10^6$.

\begin{figure}
\centering
\includegraphics[width= 3.0 in]{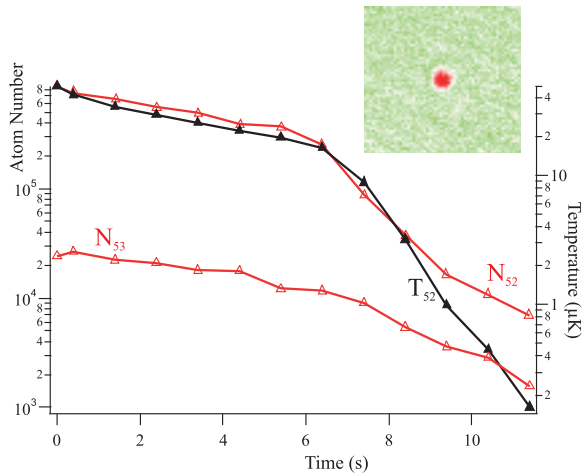}
\caption{\setlength{\baselineskip}{6pt} {\protect\scriptsize (Color online) Time evolution of $^{52}$Cr temperature and atom number of both
 isotopes. t=0 corresponds to the end of the IR trap loading. The crossed dipole trap is fully loaded at t=7 s. The evaporation ends
 at t=12 s.  An in-situ absorption image of the degenerate fermionic cloud is shown.}}
\label{Evap}
\end{figure}

\begin{figure}
\centering
\includegraphics[width= 3.0 in]{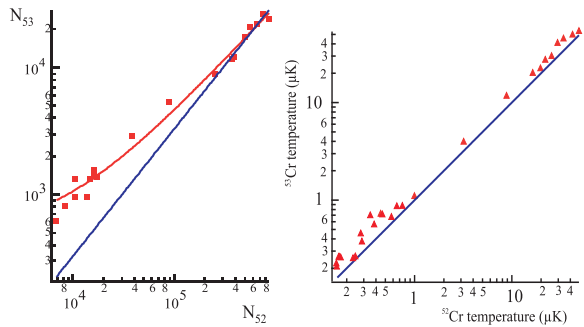}
\caption{\setlength{\baselineskip}{6pt} {\protect\scriptsize (Color online) Comparison between the two isotopes during evaporation.
Left: Atom numbers showing the smaller losses for fermions at the end of evaporation. The (red) curve is a guide for the eye, the straight (blue)
line is a linear fit to the initial trend.
Right: Temperatures. The temperature of the fermion remains about $20\%$ higher than that of the boson during the whole sequence. The straight
(blue) line corresponds to identical temperatures.}}
\label{Temperatures}
\end{figure}

Once the atoms are optically pumped to their absolute ground state (respectively $S=3,m_S=-3$ and $F=9/2,m_F=-9/2$),
the crossed dipole trap is implemented by transferring 80$\%$ of IR power
to a vertical beam, in 9 s. The total IR power is then reduced to 1 W (starting at t=5 s) in 7 s. Figure \ref{Evap} shows the
evolution of the Boson temperature and atom numbers for both isotopes. Thermalization between the two isotopes is relatively good during
the whole evaporation process as shown by Fig.\ref{Temperatures}.
Consequently, the interspecies cross section $\sigma_{BF}$ has to be relatively large. On the other
hand, thermalization is not perfect: the fermion temperature is measured to be about $20\%$ higher than the one of the
boson. We thus infer that $\sigma_{BF}$ is smaller than the bosonic cross section $\sigma_{BB}$. At the end of the ramp, a $^{52}$Cr BEC is
obtained with typically $8.10^3$ atoms, while the number of $^{53}$Cr atoms ranges between 500 and 1000.

The power of the IR laser is then rapidly ramped up (to 5 W) to obtain a tighter trap and freeze evaporation. Trap frequencies
$\omega_{x,y,z}=2\pi\times(430,510,350)$ Hz are measured through parametric excitation, with $5\%$ uncertainty.
In this trap, we obtain that the critical temperature for BEC is
$T_c=380$ nK for $10^4$ atoms, while the Fermi temperature is $T_f=370$ nK for $10^3$ atoms. The experimental temperatures are obtained by fitting
the velocity distributions imaged after a free fall, by a bimodal distribution for $^{52}$Cr, and by either \cite{NoteDeMarco} a Boltzmann or a Fermi-Dirac distribution
for $^{53}$Cr. We obtain $T_{52}= (180\pm20)$ nK and $T_{53}=(220\pm20)$ nK. We therefore obtain $T_{53}/T_f=0.6$.

In addition to report the production of $^{53}$Cr degenerate gases, one of the main points of this
paper is to explain why our strategy to reach quantum degeneracy is efficient despite the
small atom number before evaporation. For that we develop the following
theoretical model, based on the
scaling laws for evaporation first introduced in \cite{Ohara}. Our
main assumption is that polarized $^{53}$Cr atoms only collide with $^{52}$Cr atoms. This assumption
is first motivated by the fact that the centrifugal barrier is of the order of 1 mK for the $p-$wave
channel, much larger than the initial temperature of the cloud. As a consequence, polarized Fermions
can be considered not to collide with each other through Van der Walls interactions. In addition, the expected scattering cross-section
between identical Fermions due to DDIs, estimated to be $9.8\times10^{-18}$ m$^{-2}$ using first order Born
approximation \cite{Giovanazzi}, is small enough that dipolar collisions may be neglected given the density. Therefore
one can consider that fermions only collide with most abundant $^{52}$Cr atoms. Due to large differences in atom numbers, we assume as well
that evaporative losses of bosons solely come from collisions with bosons.

Given these assumptions, the rate equations for evaporation are given by:

\begin{align}
\label{evolNT}
\frac{dN_B}{dt}& =  -\sigma_{BB} n_B \bar{V}f(\eta_B) N_B - \Gamma N_B \\
\frac{dN_F}{dt} & =  -\sigma_{BF} n_B \bar{V}f(\eta_F) N_F - \Gamma N_F \nonumber \\
\frac{d(3 N_B T_B)}{dt} & =   -\sigma_{BB} n_B \bar{V}f(\eta_B) (\eta_B +1) N_B T_B - \Gamma (3 N_B T_B) \nonumber \\
\frac{d(3 N_F T_F)}{dt} & =  -\sigma_{BF} n_B \bar{V}f(\eta_F) (\eta_F +1) N_F T_F - \Gamma (3 N_F T_F)  \nonumber
\end{align}

where $i=B,F$ stands for Boson of Fermion, $N_i$ is an atom number, $n_i$ a density, $\bar{V}=(16 k_B T/\pi m)^{1/2}$ is the mean of the magnitude
of the relative velocity, $\Gamma$ is
the one body loss coefficient, independently measured
to be $\Gamma = .1$ s$^{-1}$ in our experiment, $\eta$ is the ratio between the trap depth and the thermal energy,
and $f(\eta)=2(\eta-4)e^{-\eta}$ \cite{Ohara,Doyle}. These equations assume thermal equilibrium. As noticed above the fermion temperature
 is slightly higher than the boson temperature, which signals a slight departure from thermal
equilibrium. To interpret our data, we therefore assume thermal equilibrium for each gas, at a
temperature which slightly depends on the isotope.

The first term in the right-hand side of each
equation in (\ref{evolNT}) describes evaporation. One major difficulty in applying this model
to quantitatively describe evaporation is that the rate of evaporation depends exponentially on $\eta$,
 $via$ the exponential term in $f$. It is usually difficult to precisely measure the trap depth in one experiment
(e.g. due to uncertainties in estimating waists of laser beams in situ). However, the trap depth is almost identical for the two isotopes
(isotopic shifts being much smaller than the detunings of the IR laser to all
optical transitions). In our analysis, we therefore strongly reduce the sensitivity to the trap depth
by comparing the rate of evaporation of fermions to the one of bosons. The first two equations of (\ref{evolNT}) give:

\begin{equation}
\frac{\log_e\left(\frac{N_B(t_2)}{N_B(t_1)}  \right)+
 \Gamma (t_2-t_1)}{\log_e\left(\frac{N_F(t_2)}{N_F(t_1)}  \right)+
  \Gamma (t_2-t_1)}=\frac{\sigma_{BB}}{\sigma_{BF}} \times \frac{f(\eta_B)}{f(\eta_F)}
  \label{IntegrateN}
\end{equation}

As seen from eq.(\ref{IntegrateN}), the sensitivity to trap depth is not completely suppressed by the
relative measurement, because the temperature of both clouds are slightly different, leading to
values of $\eta_{B}$ and $\eta_F$ differing by up to 20 percent. We therefore first estimate the
value of $\eta_B$ by comparing the measured loss rate of bosons and cooling rate. From the first and third eqs in (\ref{evolNT}) we infer:

\begin{equation}
\log_e\left(\frac{T_B(t_2)}{T_B(t_1)}\right)=\frac{2-\eta}{3} \left[\log_e\left(\frac{N_B(t1)}{N_B(t_2)}\right) - \Gamma(t_2-t_1)\right]
\label{IntegrateT}
\end{equation}

The analysis of our data using eq.(\ref{IntegrateT}) leads to an experimental estimate of $\eta_B=6.5 \pm 0.5$, in relatively good
agreement with the ratio of the trap depth to temperature $\approx 7$ which we infer from calibrating
the trap depth. We thus infer $f(\eta_{B})/f(\eta_{F})=2.0\pm0.5$

\begin{figure}
\centering
\includegraphics[width= 3.0 in]{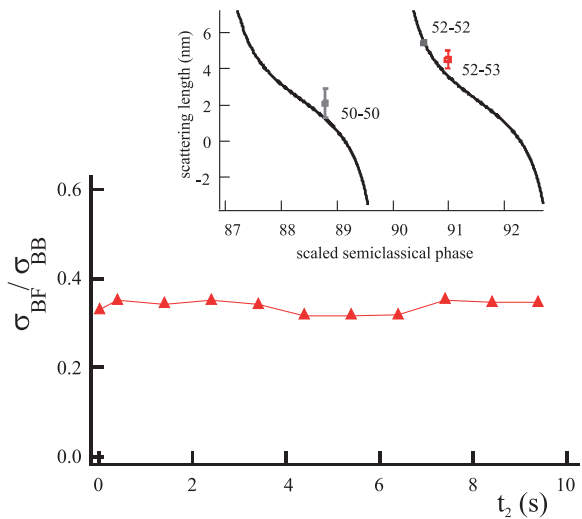}
\caption{\setlength{\baselineskip}{6pt} {\protect\scriptsize (Color online) Experimental measurements of the ratio of the boson-boson
and inter-isotope cross sections obtained using eq.(\ref{IntegrateN}), with $t_1=10$ s (see text). Inset: scattering lengths of Cr,
including $a_{52-53}$ (this work), which is in good agreement with predictions based on mass-scaling.
The solid line is the predicted scattering length using mass scaling of semi-classical phase \cite{MassScaling}.}}
\label{SigmaRatio}
\end{figure}

We now use our experimental data, to estimate from eq.(\ref{IntegrateN}) the value for the cross-section
$\sigma_{BF}$. We can use different portions of the experimental data to measure the ratio $\sigma_{BF} / \sigma_{BB}$
as shown in Fig.\ref{SigmaRatio}. In practice, this is for example done by spanning the times $t_1$ and
$t_2$ over the range of experimental times which were investigated in our experimental data.
Experimental values of $\sigma_{BF} / \sigma_{BB}$ all lie in the interval $0.35\pm0.05$ irrespective of the choice of $t_1$ and $t_2$.
This indicates that $\sigma_{BF} / \sigma_{BB}$ is insensitive to temperature to within
signal-to-noise ratio. This is in good agreement with the assumptions that atoms collide mostly through
$s-$wave and short-range interactions. This is also in good agreement with theoretical predictions
\cite{Simoniprivate}.

An outcome of our analysis is the first measurement of the boson-fermion scattering length $a_{BF}$ in the
$S=6$ electronic molecular potential (both atoms being in the stretched state of lowest energy). Indeed,
$\sigma_{BF} = 4 \pi a_{BF}^2$ describe collisions between (distinguishable) bosons and fermions, while the value of
$\sigma_{BB} = 8 \pi a_{BB}^2$, which describes collisions between (undistinguishable) bosons, has been measured with accuracy \cite{PRAreldip}.
In Fig.\ref{SigmaRatio} we report the known value of $a_6$ for $^{52}$Cr, the less-well
known value for $^{50}$Cr \cite{PfauFeshbach}, as well as our newly measured value for $|a_{BF}|=85 (\pm 10 )$ $a_{B}$. The
good agreement with theoretical predictions based on mass-scaling \cite{MassScaling} indicates that $a_{BF}>0$.

Our analysis therefore confirms that $\sigma_{BF} < \sigma_{BB}$. As evaporation is optimized to be
achieved as fast as possible for the boson, it is not surprising that the Fermi cloud lags slightly
behind in terms of temperature. This analysis shows that
our strategy to cool fermions is particularly efficient because (i) $\sigma_{BF}$ is sufficiently
large to (almost) insure inter-isotope thermal equilibrium; (ii) $\sigma_{BF}$ is small enough to
reduce evaporative losses of fermions, which leads to a gradual increase of the ratio of the number of fermions
to the number of bosons as evaporation proceeds (see Fig.\ref{Temperatures}). This increase is
essential for the positive outcome of our experiment.

In conclusion, we have produced a $^{53}$Cr degenrate Fermi gas, with up to $10^3$ atoms, together with a BEC of $10^4$ $^{52}$Cr atoms.
This boson-fermion degenerate mixture might have peculiar properties, due to the strong imbalance in atom numbers. The spatial mode
of the Fermi sea may be deformed by strong repulsive interaction with the BEC, which is of the order of the Fermi energy. Deformation of the Fermi
surface may arise as well. These effects should be amplified close to an inter-isotope Feshbach resonance.

We plan to reveal the dipolar nature of the Cr Fermi sea by studying Pauli paramagnetism  at low magnetic field. Quantum statistics is expected
to lead then to a very different picture than the one we obtained for $^{52}$Cr \cite{ThermoSpin3}. In a 3D optical lattice, $^{53}$Cr should
provide a good platform to engineer new magnetism associated to XYZ spin models \cite{Carr},
complementary to recent studies with $^{52}$Cr \cite{DSE}.

Acknowledgements: LPL is Unit\'e Mixte (UMR 7538) of CNRS and
of Universit\'e Paris 13, Sorbonne Paris Cit\'e. We thank A. Simoni for exchanges about mass scaling.
We acknowledge financial support from Conseil R\'{e}%
gional d'Ile-de-France under DIM Nano-K / IFRAF, CNRS, and from Minist\`{e}re de
l'Enseignement Sup\'{e}rieur et de la Recherche within CPER Contract.


\begin{thebibliography}{199}

\bibitem{CrBEC} A. Griesmaier, J. Werner, S. Hensler, J. Stuhler, and T. Pfau, \textit{Phys. Rev. Lett.} \textbf{94}, 160401 (2005)

\bibitem{CrBEC2} T. Koch, T. Lahaye, J. Metz, B. Fröhlich, A. Griesmaier, and T. Pfau, \textit{Nature Physics} \textbf{4}, 218 (2008)
T. Lahaye, J. Metz, B. Fröhlich, T. Koch, M. Meister, A. Griesmaier, T. Pfau, H. Saito, Y. Kawaguchi, and M. Ueda, \textit{Phys. Rev. Lett.} \textbf{101}, 080401 (2008);
G. Bismut, B. Pasquiou, E. Mar\'echal, P. Pedri, L. Vernac, O. Gorceix, and B. Laburthe-Tolra, \textit{Phys. Rev. Lett.} \textbf{105}, 040404 (2010);
G. Bismut, B. Laburthe-Tolra, E. Mar\'echal, P. Pedri, O. Gorceix, and L. Vernac, \textit{Phys. Rev. Lett.} \textbf{109}, 155302 (2012)

\bibitem{DyBEC} M. Lu, N.Q. Burdick, S.H. Youn, and B.L. Lev,  \textit{Phys. Rev. Lett.} \textbf{107}, 190401 (2011)

\bibitem{ErBEC} K. Aikawa, A. Frisch, M. Mark, S. Baier, A. Rietzler, R. Grimm, and F. Ferlaino, \textit{Phys. Rev. Lett.} \textbf{108}, 210401 (2012)

\bibitem{BaranovZoller} M. A. Baranov, M. Dalmonte, G. Pupillo, and P. Zoller \textit{Chem. Rev.} \textbf{112}, 5012 (2012)

\bibitem{pwave} M.A. Baranov, M.S. Marenko, V.S. Rychkov, and G.V. Shlyapnikov, \textit{Phys. Rev. A} \textbf{66}, 013606 (2002)

\bibitem{pxpy} J. Levinsen, N. R. Cooper, and G. V. Shlyapnikov \textit{Phys. Rev. A} \textbf{84}, 013603 (2011)

\bibitem{BuchlerPRL} D. Peter, S. Muller, S. Wessel, and H.P. Buchler, \textit{Phys. Rev. Lett.} \textbf{109}, 025303 (2012)

\bibitem{Carr} M. L. Wall, K. Maeda, and L. D. Carr, arXiv:1410.4226 (2014)

\bibitem{BlochZoller} A.W. Glaetzle, M. Dalmonte, R. Nath, C. Gross, I. Bloch, and Peter Zoller arXiv:1410.3388 (2014)

\bibitem{Buchler} D. Peter, N.Y. Yao, N. Lang, S.D. Huber, M.D. Lukin, and H.P. Buchler, arXiv:1410.5667 (2014)

\bibitem{DyFS} M. Lu, N.Q. Burdick, and B.L. Lev,	\textit{Phys. Rev. Lett.} \textbf{108}, 215301 (2012)

\bibitem{ErFS} K. Aikawa, A. Frisch, M. Mark, S. Baier, R. Grimm, and F. Ferlaino \textit{Phys. Rev. Lett.} \textbf{112}, 010404

\bibitem{DefFermiSurface}K. Aikawa, S. Baier, A. Frisch, M. Mark, C. Ravensbergen, F. Ferlaino  \textit{Science} \textbf{345}, 1484 (2014)

\bibitem{FeshbachEr} A. Frisch, M. Mark, K. Aikawa, F. Ferlaino, J. L. Bohn, C. Makrides, A. Petrov, and S. Kotochigova,
\textit{Nature} \textbf{507}, 475-479 (2014)

\bibitem{FeshbachDy} K. Baumann, N.Q. Burdick, M. Lu, and B.L. Lev, \textit{Phys. Rev. A} \textbf{89}, 020701(R) (2014)

\bibitem{MOT5253} R. Chicireanu, A. Pouderous, R. Barb\'e, B. Laburthe-Tolra, E. Mar\'echal, L. Vernac, J.-C. Keller and O. Gorceix, \textit{Phys. Rev. A} textbf{73}, 053406 (2006).

\bibitem{MassScaling} G. F. Gribakin and V. V. Flambaum \textit{Phys. Rev. A} \textbf{48}, 546 (1993);
V. V. Flambaum, G. F. Gribakin, and C. Harabati \textit{Phys. Rev. A} \textbf{59}, 1998 (1999)

\bibitem{FirstFS} B. DeMarco and D. S. Jin, \textit{Science} \textbf{285} 1703  (1999)

\bibitem{Li67FS} A.G. Truscott, K.E. Strecker, W.I. McAlexander, G.B. Patridge, and R.G. Hulet, \textit{Science} \textbf{291}, 2570 (2001);
F. Schreck, L. Khaykovich, K.L. Corwin, G. Ferrari, T. Bourdel, J. Cubizolles, and C. Salomon, \textit{Phys. Rev. Lett.} \textbf{87}, 080403 (2001)

\bibitem{CrBECus} Q. Beaufils, R. Chicireanu, T. Zanon, B. Laburthe-Tolra,
E. Mar\'echal, L. Vernac, J.-C. Keller, and O. Gorceix, \textit{Phys. Rev. A} \textbf{77}, 061601 (2008)


\bibitem{Note} Analysis of the loading optimization, as well as technical details of the experiment, and spectroscopic data of repumping lines,
will be presented in a forthcoming paper.

\bibitem{CrBECAppPhys} G. Bismut, B. Pasquiou, D. Ciampini, B. Laburthe-Tolra, E. Mar\'echal, L. Vernac, and O. Gorceix, \textit{Appl. Phys. B} \textbf{102} 1-9 (2011)


\bibitem{NoteDeMarco} For $T/T_F=0.6$, it is expected that fitting the Fermi Dirac distribution by the Boltzmann statistics leads to the same
value of temperature within  $3\%$ (B. DeMarco, PhD thesis, University of Colorado (2001)).


\bibitem{Ohara} K. M. O'Hara, M. E. Gehm, S. R. Granade, and J. E. Thomas, \textit{Phys. Rev. A} \textbf{64}, 051403(R)

\bibitem{Giovanazzi} S. Hensler, J. Werner, A. Griesmaier, P.O. Schmidt, A. Görlitz, T. Pfau, S. Giovanazzi and K. Rzazewski \textit{Appl. Phys. B} \textbf{77}, 765 (2003)

\bibitem{Doyle} R. de Carvalho and J. Doyle, \textit{Phys. Rev. A} \textbf{70}, 053409 (2004)

\bibitem{Simoniprivate} A. Simoni, private communication

\bibitem{PRAreldip} B. Pasquiou, G. Bismut, Q. Beaufils, A. Crubellier, E. Mar\'echal, P. Pedri, L. Vernac, O. Gorceix, and B. Laburthe-Tolra
\textit{Phys. Rev. A} \textbf{81}, 042716 (2010)

\bibitem{PfauFeshbach} J. Werner, A. Griesmaier, S. Hensler, J. Stuhler, T. Pfau, A. Simoni and E. Tiesinga, \textit{Phys. Rev. Lett.} \textbf{94}, 183201 (2005)

\bibitem{ThermoSpin3} B. Pasquiou, E. Mar\'echal, L. Vernac, O. Gorceix, and B. Laburthe-Tolra, \textit{Phys. Rev. Lett.} \textbf{108}, 045307 (2012)

\bibitem{DSE}  A. de Paz, A. Sharma, A. Chotia, E. Mar\'echal, J. H. Huckans, P. Pedri, L. Santos, O. Gorceix, L. Vernac, and B. Laburthe-Tolra,
Phys. Rev. Lett. \textbf{111}, 185305 (2013)

\end{thebibliography}
\end{document}